\documentstyle[12pt]{article}

\begin{document}

\font\ninerm = cmr9



\def\footnoterule{\kern-3pt \hrule width \hsize \kern2.5pt}

\pagestyle{empty}

\begin{flushright}
hep-th/0105120 \\
$~$ \\
\end{flushright}

\vskip 0.5 cm

\begin{center}

{\large\bf Coproduct and star product in field theories\\
on Lie-algebra non-commutative space-times}

\end{center}

\vskip 1.5 cm

\begin{center}

{\bf Giovanni AMELINO-CAMELIA} and {\bf Michele ARZANO}\\

\end{center}

\begin{center}

{\it Dipart.~Fisica,
Univ.~Roma ``La Sapienza'',
P.le Moro 2, 00185 Roma, Italy}
\end{center}

\vspace{1cm}

\begin{center}

{\bf ABSTRACT}

\end{center}

{\leftskip=0.6in \rightskip=0.6in

We propose a new approach to field theory on $\kappa$-Minkowski
non-commutative space-time, a popular example of
Lie-algebra space-time. Our proposal is essentially
based on the introduction of a star product, a technique which is
proving to be very fruitful in analogous studies of canonical
non-commutative space-times, such as the ones recently found
to play a role in the description of certain string-theory backgrounds.
We find to be incorrect the expectation, previously reported in
the literature, that the lack of symmetry of the $\kappa$-Poincar\'{e}
coproduct should lead to interaction vertices that are not
symmetric under exchanges of the momenta of identical particles
entering the relevant processes.
We show that in $\kappa$-Minkowski the coproduct and the
star product must indeed treat
momenta in a non-symmetric way, but the
overall structure of interaction vertices is symmetric
under exchange of identical particles.
We also show that in $\kappa$-Minkowski field theories
it is convenient to introduce the concepts
of ``planar" and ``non-planar" Feynman loop-diagrams, again in close
analogy with the corresponding concepts previously introduced in
the study of field theories in canonical
non-commutative space-times.}

\newpage



\baselineskip 20pt plus .5pt minus .5pt

\pagenumbering{arabic}

\pagestyle{plain}


Non-commutative geometry is being used more and more extensively
in attempts to unify general relativity and quantum mechanics.
Some ``quantum-gravity"
approaches explore the possibility that non-commutative geometry
might provide the correct fundamental description of space-time,
while in other approaches non-commutative geometry
turns out to play a role at the level of the effective theories
that describe certain aspects of quantum-gravity.

Two simple examples~\cite{starwess}
are ``canonical non-commutative space-times" ($\mu,\nu,\beta = 0,1,2,3$)
\begin{equation}
[x_\mu,x_\nu] = i \theta_{\mu,\nu}
\label{canodef}
\end{equation}
and ``Lie-algebra non-commutative space-times"
\begin{equation}
[x_\mu,x_\nu] = i C^\beta_{\mu,\nu} x_\beta ~.
\label{liedef}
\end{equation}
The canonical type (\ref{canodef}) was originally proposed~\cite{dopl}
in the context of attempts to develop a new fundamental picture
of space-time. More recently, (\ref{canodef}) is proving useful in the
description of string theory in certain ``$B$" backgrounds
(see, {\it e.g.}, Refs.~\cite{seibwitt,suss,chong}).
String theory in these backgrounds admits description (in the sense
of effective theories) in terms of a field theory in the non-commutative
space-times (\ref{canodef}), with the tensor $\theta_{\mu,\nu}$
reflecting the properties of the specific background.

Among the Lie-algebra type (\ref{liedef}) space-times,
one of the most studied
is $\kappa$-Minkowski~\cite{majrue,kpoinap}
space-time ($l,m = 1,2,3$)
\begin{equation}
[x_m,t] = {i \over \kappa} x_m ~,~~~~[x_m, x_l] = 0 ~.
\label{kmindef}
\end{equation}
One of us recently proposed~\cite{dsr1dsr2} a new path toward
quantum gravity which takes as starting point $\kappa$-Minkowski.
It was shown in Ref.~\cite{dsr1dsr2} that the Hopf algebra
that characterizes the symmetries of $\kappa$-Minkowski,
one of the $\kappa$-Poincar\'{e} algebras~\cite{kpoinap,kpoinori},
can be used to introduce the Planck length (here identified, up to
a sign and perhaps a numerical factor of order 1, with $1/\kappa$)
directly at the level of the Relativity postulates.
This could provide~\cite{dsr1dsr2}
the starting point for a path toward
quantum gravity based on a, still being sought,
non-flat (and dynamical) generalization of the $\kappa$-Minkowski
non-commutative space-time.

The possible role in quantum gravity of non-commutative geometry,
and particularly of the examples (\ref{canodef}) and (\ref{liedef}),
has generated strong interest in the
construction of field theories in these types of space-times.
In this respect the canonical type confronts us with less
severe technical challenges. Procedures based on the
star product~\cite{starwess} can be applied straightforwardly
to space-times of type (\ref{canodef}), basically as a result of the fact
that the product of two wave exponentials, $e^{i p_\mu x_\mu}$
and $e^{i k_\nu x_\nu}$, in these geometries has very simple properties:
\begin{equation}
e^{i p_\mu x_\mu} e^{i k_\nu x_\nu} =
e^{i p_\mu \theta_{\mu,\nu} k_\nu} e^{i (p +k)_\mu x_\mu}
~,
\label{expprodcano}
\end{equation}
{\it i.e.} the Fourier parameters $p_\mu$ and $k_\mu$ combine just as
usual, $(p +k)_\mu$, with the only new ingredient of an overall phase
factor that depends on $\theta_{\mu,\nu}$. This simplicity renders
possible the development of field theories
in which the tree-level propagator is undeformed and the only new
ingredients are factors of the type $e^{i p_\mu \theta_{\mu,\nu} k_\nu}$
in the interaction vertices. These properties follow from the analysis
of products of fields, which are characterized by
formulas of the type
\begin{eqnarray}
\Phi(x) \Psi(x) & = & \frac{1}{(2\pi)^4} \int
 d^4p \, d^4k~e^{i p   x } ~ \tilde{\Phi}(p) e^{i p   k } \tilde{\Psi}(k)
\label{starcano}\\ & = & \frac{1}{(2\pi)^4} \int
 d^4p \, d^4k~e^{i p_\mu \theta_{\mu,\nu} k_\nu}~
e^{i (k+p)   x } ~ \tilde{\Phi}(p) ~ \tilde{\Psi}(k)
~.
\nonumber
\end{eqnarray}

It is easy to see that in space-times of Lie-algebra type, and in particular
in $\kappa$-Minkowski, the construction of field theories is
less straightforward. A key point is that the type of simple deformation
encoded in (\ref{expprodcano}) is not sufficient.
A more complicated rule for
the product of two wave exponentials is clearly needed, basically to
reflect the structure of the coproduct in the $\kappa$-Poincar\'{e}
algebras~\cite{kpoinap}. The rule for the product of two wave
exponentials treats in profoundly\footnote{Here we are emphasizing
that in $\kappa$-Minkowski the $k \rightarrow p,~ p \rightarrow k$
symmetry of the product of wave exponentials is lost in a much more
significant way than in (\ref{expprodcano}). [In (\ref{expprodcano})
the exchange
of $k$ and $p$ only changes the sign of the constant phase factor.]}
non-symmetric way the two Fourier parameters.
This has been cause of concern with respect to the applicability
of $\kappa$-Minkowski in physics, since it
appeared~\cite{lukiFT,gacmaj,lukistar}
inevitable to obtain cross sections which do not
treat symmetrically pairs of identical incoming particles.
The most significant result here being reported is that
no such paradoxical predictions are found when field theory
in $\kappa$-Minkowski space-time is constructed consistently.

Our first task is the one of giving a consistent
description of the product of two fields in $\kappa$-Minkowski space-time
in terms of deformed rules for the product of commuting fields.
This is the basic task of
the star-product procedure: the product of two operator-valued
functions is described through the properties of an associated deformed
product of ordinary functions.
We procede in analogy with the strategy
(\ref{starcano}) which is proving successful in the study of
canonical non-commutative spacetimes, {\it i.e.} we want to introduce
as auxiliary commuting variables some Fourier parameters.
The proper formulation of the Fourier transform
in $\kappa$-Minkowski space-time has been discussed in previous
mathematical-physics studies~\cite{gacmaj,lukistar,majoek};
it is based on the ``ordered exponential" $:e^{i p_\mu x_\mu}:$
defined by\footnote{There is of course an equally valid alternative
ordering prescription in which the time-dependent exponential is
placed to the
left~\cite{lukistar} (while we are here choosing the
convention with the time-dependent exponential to the right).}
\begin{equation}
:e^{i p_\mu x_\mu}: \equiv e^{i p_m x_m}   e^{i p_0 x_0}
~.
\label{order}
\end{equation}
While, as the reader can easily verify,
wave exponentials of the type $e^{i p_\mu x_\mu}$ do not combine
in a simple way in $\kappa$-Minkowski space-time,
for wave exponentials of the type $:e^{i p_\mu x_\mu}:$
one finds~\cite{lukistar} (using the Campbell-Baker-Hausdorff formula)
\begin{equation}
(:e^{i p_\mu x_\mu}:) (:e^{i k_\nu x_\nu}:) =
:e^{i (p \dot{+} k)_\mu x_\mu}:
~,
\label{expprodlie}
\end{equation}
where the notation ``$\dot{+}$" has been here introduced to denote
the deformed addition rule (no sum on repeated indices)
\begin{equation}
p_\mu \dot{+} k_\mu \equiv \delta_{\mu,0}(p_0+k_0) + (1-\delta_{\mu,0})
(p_\mu +e^{p_0/\kappa} k_\mu) ~,
\label{coprod}
\end{equation}
{\it i.e.} the energy\footnote{We are here taking the liberty to
denominate ``energy" the Fourier parameter in the 0-th direction  (and
similarly for the other three Fourier parameters we use ``3-momentum").
This terminology may appear unjustified in the present context,  but it is
actually meaningful in light of the results on $\kappa$-Minkowski  reported
in Ref.~\cite{gacmaj}.}  addition is undeformed while 3-momenta  are added
according to $\vec{p}+e^{\lambda p_0}\vec{k}$.  This lack of symmetry under
$\vec{p}$,$\vec{k}$ exchange is the  one we announced in the opening
remarks, and readers familiar  with the $\kappa$-Poincar\'{e} research
programme will recognize that  (\ref{coprod}) is just the rule for
energy-momentum coproduct.    As mentioned, the proper
formulation~\cite{gacmaj,lukistar,majoek}  of the Fourier transform in
$\kappa$-Minkowski space-time  is based on the ordered exponentials
$:e^{ip_\mu x_\mu}:$
\begin{equation}
\Phi(x)=\frac{1}{(2\pi)^2} \int d^4p ~ :e^{i p   x }: ~
  \tilde{\Phi}(p) ~.
\label{wessfourier}
\end{equation}
The canonical property of Fourier theory is then obtained~\cite{gacmaj}
using the proper concept of partial derivative in $\kappa$-Minkowski:
\begin{equation}
{\partial \over \partial x_m} :e^{i p   x }:
= : {\partial \over \partial x_m} e^{i p   x }: ~,~~~~
{\partial \over \partial x_0} :e^{i p   x }:
= \kappa : \left( e^{i p   x }
- e^{i p   (x + \Delta x_\kappa)} \right):
~,
\label{partialderiv}
\end{equation}
where $(\Delta x_\kappa)_\mu \equiv - \delta_{\mu,0}/\kappa$.

Using (\ref{expprodlie}) one easily finds that
\begin{equation}
\Phi(x) \Psi(x) = \frac{1}{(2\pi)^4} \int
 d^4p \, d^4k ~ :e^{i (p \dot{+} k)   x}:
~ \tilde{\Phi}(p) ~ \tilde{\Psi}(k)
~.
\label{starlie}
\end{equation}
Having established this simple property of the product of two fields in
$\kappa$-Minkowski space-time we can now easily write a partition function
(generating functional for Green functions) in energy-momentum space.
This will allow us to explore whether the lack of symmetry of the
coproduct leads to nonsensical results, as suspected in previous related
studies~\cite{lukiFT,gacmaj,lukistar}, or instead an acceptable physical
picture does emerge. We illustrate our line of analysis in the context of
a scalar theory with quartic interaction (``$\lambda \Phi^4$ theory").
We start with a partition function in the non-commutative space-time
\begin{equation}
Z[J(x)]=\int {\mathcal D}[\phi]\,e^{ i \int
  d^4x \, [\frac{1}{2} \partial^{\mu}\phi(x)\partial_{\mu}\phi(x)
-\frac{m^2}{2} \phi^2(x) - \frac{\lambda}{24}\phi^4(x)
+\frac{1}{2} J(x)\phi(x)+ \frac{1}{2} \phi(x)J(x)]}~.
\label{zeta1st}
\end{equation}
The observation (\ref{starlie}) allows to rewrite this partition
function in energy-momentum space.
We omit the tedious steps of this derivation, but we note
here some useful formulas which reflect the type of care
required by the lack of symmetry of the coproduct. Introducing
\begin{equation}\label{deltax}
\delta^{(4)}(k)=\frac{1}{(2\pi)^4}\int d^4 x \, :e^{i k   x}: ~,
\end{equation}
which holds in $\kappa$-Minkowski space-time~\cite{lukistar},  the required
manipulations of the partition function will of course  lead to the
emergence of terms going like $\delta^{(4)}(p \dot{+} k)$.
The lack of symmetry of $p \dot{+} k$ then requires some care;
the relevant formulas are
\begin{equation}
\int d^4k \, \delta^{(4)}(k\dot{+}p)f(k)= \mu(p_0) \, f(\dot{-}p)
~,
\label{deltaone}
\end{equation}
\begin{equation}
\int d^4k \, \delta^{(4)}(p\dot{+}k)f(k)=e^{-3\lambda p_0}
\, \mu(p_0) \, f(\dot{-}p) ~,
\label{deltatwo}
\end{equation}
where we introduced a function $\mu(p_0)$, possibly reflecting  the
properties of a non-trivial measure of integration   over
$\kappa$-energy-momentum space   (which, in the sense reflected by
(\ref{coprod}), is not flat)  and we also introduced the notation
$\dot{-}p$
\begin{equation}
(\dot{-} p)_\mu \equiv \delta_{\mu,0}(-p_0) + (1-\delta_{\mu,0})
(-e^{-p_0/\kappa} p_\mu) ~.
\label{inverse}
\end{equation}
Readers familiar with the $\kappa$-Poincar\'{e} research programme will
recognize  (\ref{inverse}) as the rule for the ``antipode"
(in fact $p \dot{+} (\dot{-} p) =0$). On the function
$\mu(p_0)$ we will only observe and use the fact that  it can depend on
energy-momentum only through the 0-th component  (energy).
As discussed in detail in Ref.~\cite{gacmaj}, it is clear that  the measure
for integration over $\kappa$-energy-momentum   space should depend only on
energy. There are various candidates  for this measure, but for our
analysis this ambiguity could only affect the  form of the function
$\mu(p_0)$.  Since we focus here on the relation between the non-symmetric
coproduct of $\kappa$-Poincar\'{e} and the structure of the conservation  rules
that characterize Green functions in field theory on $\kappa$-Minkowski, we
can postpone the investigation of the measure ambiguity  to future studies.
Concerns about the conservation rules have  been the most serious obstacle
for the construction of physical  theories based on $\kappa$-Minkowski, and
we shall show that these  concerns can be straightforwardly addressed
within our approach,  independently of the form of $\mu(p_0)$. We shall
therefore keep  track of factors of the type $\mu(p_0)$, but never assume
anything  about the form of the function $\mu(p_0)$.  While the existence
of alternative choices~\cite{gacmaj} of measure  (leading to different
forms of our $\mu(p_0)$) is  an ``aesthetic concern" for the
$\kappa$-Minkowski research programme  (all the candidates identified in
Ref.~\cite{gacmaj} are acceptable,  but one would like to have a theory
predicting the measure),  the possibility that Green functions might be
characterized by conservation  rules that do not respect symmetry under
exchange of the momenta of  identical incoming (outgoing)
particles represents a ``quantitative
concern" for the  applicability of $\kappa$-Minkowski in physics. As
anticipated,  it is on this second, more alarming, problem which we focus
here our attention.

Our first objective is to examine the structure of the tree-level propagator.
For this first result we can of course switch off the coupling $\lambda$.
Using (\ref{partialderiv}), (\ref{deltaone}) and (\ref{deltatwo}),  one
obtains from (\ref{zeta1st})
\begin{eqnarray}
Z^0[J(k)] \! \equiv \! \{Z[J(k)]\}_{\lambda=0} \! = \!
\int \!\! {\mathcal D}[\phi]\,e^{\frac{i}{2}\int \! d^4k \, \mu(k_0)
[\phi(\dot{-}k) [{\mathcal C}_{\kappa}(k)-m^2] \phi(k)
+J(k)\phi(\dot{-}k)+\phi(k)J(\dot{-}k)]}
\label{zeta2nd}
\, ,
\end{eqnarray}
where ${\mathcal C}_{\kappa}$ is the $\kappa$-Poincar\'{e} mass
casimir\footnote{We remind the reader that in the  $\kappa$-Poincar\'{e}
mass-casimir relation ${\mathcal C}_{\kappa}(k)=m^2$  the mass parameter
$m$, which here appears also in the Lagrangian,  is not to
be identified with the rest energy $E(\vec{k}=0)$.  The physical mass $M$,
which is the rest energy,  is obtained from $m$ through the
relation~\cite{kpoinap} $m^2 = \kappa^{2} sinh^2(M/(2 \kappa))$.  (Note
that however $m$ and $M$ differ only at order $1/\kappa^2$.)}
\begin{equation}
{\mathcal C}_{\kappa}(k)=\kappa^{2}(e^{k_0/\kappa}+e^{-k_0/\kappa}-2)
-\vec{k}^2 e^{-k_0/\kappa}
~.
\label{casimir}
\end{equation}
It is convenient to introduce the normalized partition function
\begin{equation}
\bar{Z}^0[J(k)] \equiv \frac{Z^0[J(k)]}{Z^0[0]}
~,
\label{normalized}
\end{equation}
and from (\ref{zeta2nd}) with simple manipulations one finds that
\begin{equation}
 \bar{Z}^0[J(k)]=exp\left(-\frac{i}{2}\int d^4k ~ \mu(k_0)
\frac{J(k)J(\dot{-}k)}{{\mathcal C}_{\kappa}(k)-m^2}\right)
~.
\label{zeta3}
\end{equation}
From Eq.~(\ref{zeta3}) one can obtain the tree-level  propagator using a
straightforward generalization  of the usual formulation adopted for field
theory in commutative  space-times:
\begin{equation}
G^{(2)}_{0}(p\,,\dot{-}p')= -\frac{\delta^2
\bar{Z}^0[J(k)]}{\delta J(\dot{-}p)\delta J(p')}\Bigg|_{J=0}
~.
\label{defprop}
\end{equation}

For the functional derivatives required by (\ref{defprop})
one also needs appropriately generalized definitions:
\begin{equation}
\frac{\delta F(f(p))}{\delta f(k)}
=\lim_{\varepsilon\rightarrow 0}\frac{1}{\varepsilon}
\left(F[f(p)+\varepsilon\delta^{(4)}(p\,\dot{+}(\dot{-}k))]-F[f(p)] \right)
~,
\label{functder1}
\end{equation}
\begin{equation}
\frac{\delta F[f(p)]}{\delta f(\dot{-}k)}
=\lim_{\varepsilon\rightarrow 0}\frac{1}{\varepsilon}
\left(F[f(p)+\varepsilon\delta^{(4)}(p\,\dot{+}k)]-F[f(p)] \right)
~.
\label{functder2}
\end{equation}

Using (\ref{deltaone}),(\ref{deltatwo}),(\ref{functder1}), (\ref{functder2})
and the property ${\mathcal C}_{\kappa}(\dot{-}p)={\mathcal C}_{\kappa}(p)$,
from (\ref{defprop}) one easily obtains the tree-level propagator:
\begin{equation}
G^{(2)}_{0}(p\,,\dot{-}p')=\frac{i}{2} \mu(-p_0)\mu(p_0)
\frac{ \delta^{(4)}(p\dot{+}(\dot{-}p'))
+ \delta^{(4)}((\dot{-}p')\dot{+}p)} {{\mathcal C}_{\kappa}(p)-m^2} ~.
\label{prop}
\end{equation}
One first point that deserves to be emphasized is the central  role played
by mass casimir ${\mathcal C}_{\kappa}$  in the structure of this
tree-level propagator.\footnote{This is  a first characteristic result of
our Lie-algebra non-commutative  space-time; in fact,   in canonical
non-commutative space-times  the tree-level propagator is governed by the
undeformed  casimir $E^2-p^2$ (but of course eventually, through loop
effects,  even in canonical non-commutative space-times  the propagator
does reflect the space-time deformation~\cite{suss}).}  This result on the
role of ${\mathcal C}_{\kappa}$ is consistent  with the expectations on the
tree-level propagator formulated  in previous preliminary
attempts~\cite{lukiFT,gacmaj} to develop   field theory in
$\kappa$-Minkowski space-time.  However, in addition to the role of
${\mathcal C}_{\kappa}$,  it is perhaps even more important to observe
that the two $\delta^{(4)}$ in (\ref{prop})  enforce the same trivial
conservation condition; in fact,
$\delta^{(4)}((\dot{-}p')\dot{+}p)
=e^{3p_0/\kappa}\delta^{(4)}(p\dot{+}(\dot{-}p'))
=e^{3p_0/\kappa}\delta^{(4)}(p-p')$.
This is a first reassuring sign that our approach  to field theory in
$\kappa$-Minkowski space-time  can provide the correct way to handle the
non-trivial  coproduct structure of $\kappa$-Poincar\'{e}: in spite of the
nonsymmetric  and nonlinear coproduct structure, our result preserves the
usual   property that energy-momentum is conserved along the tree-level
propagator.
In order to be able to investigate the properties of the
propagator  beyond tree level and in order to establish the form of the
tree-level  vertex we now push our analysis one step further,  by analyzing
the $O(\lambda)$ contributions to the Green functions.  For this we must of
course reinstate $\lambda \ne 0$, {\it i.e.}  we need to analyze
$\bar{Z}[J(k)]$ rather than $\bar{Z}^0[J(k)]$.  It turns out to be useful
to rely on the following relation between $\bar{Z}[J(k)]$ and
$\bar{Z}^0[J(k)]$, which one easily obtains with manipulations analogous to
the ones described above:
\begin{equation}
\bar{Z}[J(k)]
=e^{i\frac{\lambda}{24}
\int\ \delta^{(4)}\left(\dot{\sum}_{k_1,k_2,k_3,k_4}\right)
\prod_{j=1}^4\frac{d^4k_j}{2\pi}
\xi(k_{j,0})\frac{\delta}{\delta J(\dot{-}k_j)}}
\bar{Z}^0[J(k)]
~,
\label{intpart1}
\end{equation}
where
\begin{equation}\label{}
\xi(k_{j,0}) \equiv
2 \left(\mu(k_{j,0})+\mu(-k_{j,0})e^{\frac{3k_{j,0}}{\kappa}}\right)^{-1} ~,
\end{equation}
and we introduced a compact ordered-sum notation

\begin{equation}
\dot{\sum}_{k_1,k_2,k_3,k_4} \equiv
k_1 \dot{+} k_2 \dot{+} k_3 \dot{+} k_4
~.
\label{sumdot}
\end{equation}

Of course, for the $O(\lambda)$ contributions to the propagator and the
vertex we only need the $O(\lambda)$ approximation of $\bar{Z}[J(k)]$
\begin{equation}
\bar{Z}^{(1)}[J(k)]
=i\frac{\lambda}{24}
\int\ \delta^{(4)}\left(\dot{\sum}_{k_1,k_2,k_3,k_4}\right)
\prod_{j=1}^4\frac{d^4k_j}{2\pi}
\xi(k_{j,0})\frac{\delta}{\delta J(\dot{-}k_j)} \bar{Z}^0[J(k)]
~.
\label{zeta1stzwei}
\end{equation}
From this formula it is straightforward to obtain the $O(\lambda)$
contribution to the propagator:
\begin{eqnarray}
G^{(2)}_{\lambda}(p\,,\dot{-}p') \!\!&=&\!\!
\left(-\frac{\delta^2
\bar{Z}^1[J(k)]}{\delta J(\dot{-}p)\delta J(p')}\Bigg|_{J=0}\right)_{connected}
=
\nonumber\\
& & \!\!\!\!\!\!\!\!\!\!\!\!\!\!\!\!\!
i\frac{\lambda}{24} \int\ \prod_{j=1}^4\frac{d^4k_j}{2\pi}\xi(k_{j,0})
~ \delta^{(4)}\left(\dot{\sum}_{k_1,k_2,k_3,k_4}\right)   ~~~~~~~~~~
\label{onelooprop}\\
& & \!\!\!\!\!\!\!\!\!\!\!\!\!\!\!\!\!\!\!\!\!\!\!\!\!
\left[\frac{\delta^2
\bar{Z}^0(J(k))}{\delta J(\dot{-}p)\delta J(\dot{-}k_2)}\Bigg |_{J=0}
\frac{\delta^2 \bar{Z}^0(J(k))}{\delta J(p')\delta J(\dot{-}k_3)}
\Bigg |_{J=0}\frac{\delta^2 \bar{Z}^0(J(k))}
{\delta J(\dot{-}k_1)\delta J(\dot{-}k_4)}\Bigg |_{J=0}
\!\! + {\mathcal P}_{k_1,k_2,k_3,k_4} \right] ~,
\nonumber
\end{eqnarray}
where ${\mathcal P}_{k_1,k_2,k_3,k_4}$ denotes permutation
of the momenta $k_1,k_2,k_3,k_4$ (the term explicitly written out in the
square brackets is only one of 24 terms obtained by permutations
of $k_1,k_2,k_3,k_4$).

Let us focus on the contribution to $G^{(2)}_{\lambda}$ coming from
the first term in the square brackets of Eq.~(\ref{onelooprop}).
Using again the same observation that took us from (\ref{defprop})
to (\ref{prop}), we can rewrite this contribution as
\begin{eqnarray}
&& i\frac{\lambda}{24} \int\ \prod_{j=1}^4\frac{d^4k_j}{2\pi}\xi(k_{j,0})
~ \delta^{(4)}\left(\dot{\sum}_{k_1,k_2,k_3,k_4}\right)
\left(-\frac{i}{2}\right)^3\nonumber\\
&& ~~~~~~~~~~~~~\left(\mu(p_0)\mu(-p_0)
 \frac{1}{2} \frac{ \delta^{(4)}(p\dot{+}k_2)+
\delta^{(4)}(k_2\dot{+}p)}
{{\mathcal C}_{\kappa}(p)-m^2}\right) \label{g2lambda} \\ && ~~~~~~~~~~~~~
\left(\mu(p'_0)\mu(-p'_0)
 \frac{1}{2} \frac{ \delta^{(4)}(\dot{-}p'\dot{+}k_3)+
\delta^{(4)}(k_3\dot{+}(\dot{-}p'))}
{{\mathcal C}_{\kappa}(p')-m^2}\right)\nonumber\\ &&
~~~~~~~~~~~~~\left(\mu(k_{4,0})\mu(-k_{4,0})
 \frac{1}{2} \frac{\delta^{(4)}(k_1\dot{+}k_4)+
\delta^{(4)}(k_4\dot{+}k_1)}
{{\mathcal C}_{\kappa}(k_1)-m^2}\right) ~,
\nonumber
\end{eqnarray}
and integrating over $k_1,k_2,k_3$ we find that (\ref{g2lambda})
is proportional to
\begin{equation}
  \delta^{(4)}(\dot{-}k_4\dot{+}(\dot{-}p)\dot{+}p'\dot{+}k_4)\sim
  \delta^{(4)}(p-p') ~.
 \label{usualcons}
\end{equation}
The same observation applies to the other contributions to
$G^{(2)}_{\lambda}$ that correspond to ``planar diagrams": the 16
permutations, out of the 24 permutations encoded in ${\mathcal
P}_{k_1,k_2,k_3,k_4}$, that are such that the external momenta are attached
to consecutive internal lines ({\it i.e.} either to $k_1,k_2$ or $k_2,k_3$
or $k_3,k_4$ or $k_4,k_1$). The difference between these ``planar" diagrams
and the diagrams, which can be described as ``non-planar",
that correspond to the remaining 8 permutations, in which
instead the external lines are attached to non-consecutive lines, is
meaningful in our $\kappa$-Minkowski field theory, since the coproduct sum
$\dot{\sum}_{k_1,k_2,k_3,k_4}$ is not invariant under
permutations\footnote{Analogous comments apply to field theories in
canonical noncommutative spacetime, where indeed a completely analogous
concept of planar and non-planar diagrams has been
introduced~\cite{seibwitt,suss,chong}.}. While planar diagrams, as shown in
Eqs.~(\ref{g2lambda})-(\ref{usualcons}), provide contributions that are
proportional to the \underline{ordinary} $\delta^{(4)}(p-p')$, {\it i.e.}
correspond to ordinary conservation of energy-momentum, non-planar diagrams
introduce a new structure in the propagator. Let us consider for example
the contribution:
\begin{eqnarray}
i\frac{\lambda}{24} \int\
\prod_{j=1}^4\frac{d^4k_j}{2\pi}\xi(k_{j,0}) & & \!\!\!\!\!
\delta^{(4)}\left(\dot{\sum}_{k_1,k_2,k_3,k_4}\right)
~~~~~~~~~
\label{nonplanar1}\\
& & \!\!\!\!\!\!\!\!\!\!\!\!\!\!\!\!\!\!\!\!\!\!\!\!\!
\frac{\delta^2
\bar{Z}^0(J(k))}{\delta J(\dot{-}p)\delta J(\dot{-}k_2)}\Bigg
|_{J=0}
\frac{\delta^2 \bar{Z}^0(J(k))}{\delta J(p')\delta J(\dot{-}k_4)}
\Bigg |_{J=0}\frac{\delta^2 \bar{Z}^0(J(k))}
{\delta J(\dot{-}k_1)\delta J(\dot{-}k_3)}\Bigg |_{J=0}
\!\!  ~,
\nonumber
\end{eqnarray}
that corresponds to
\begin{eqnarray}
&& i\frac{\lambda}{24} \int\
\prod_{j=1}^4\frac{d^4k_j}{2\pi}\xi(k_{j,0})
~ \delta^{(4)}\left(\dot{\sum}_{k_1,k_2,k_3,k_4}\right)
\left(-\frac{i}{2}\right)^3\nonumber\\
&& ~~~~~~~~~~~~~\left(\mu(p_0)\mu(-p_0)
 \frac{1}{2} \frac{ \delta^{(4)}(p\dot{+}k_2)+
\delta^{(4)}(k_2\dot{+}p)}
{{\mathcal C}_{\kappa}(p)-m^2}\right) \label{g2lambdanp} \\ &&
~~~~~~~~~~~~
\left(\mu(p'_0)\mu(-p'_0)
 \frac{1}{2} \frac{ \delta^{(4)}(\dot{-}p'\dot{+}k_4)+
\delta^{(4)}(k_4\dot{+}(\dot{-}p'))}
{{\mathcal C}_{\kappa}(p')-m^2}\right)\nonumber\\ &&
~~~~~~~~~~~~~\left(\mu(k_{3,0})\mu(-k_{3,0})
 \frac{1}{2} \frac{\delta^{(4)}(k_1\dot{+}k_3)+
\delta^{(4)}(k_3\dot{+}k_1)}
{{\mathcal C}_{\kappa}(k_1)-m^2}\right) ~.
\nonumber
\end{eqnarray}
Integrating over $k_1,k_2,k_3$ we can observe that this
contribution is proportional to:
\begin{equation}
  \delta^{(4)}(\dot{-}k_3\dot{+}(\dot{-}p)\dot{+}k_3\dot{+}p')\sim
  \delta(p_0-p'_0)\,\,\delta^{(3)}(e^{\lambda p_0}\vec{k_3}-\vec{p}+\vec{k_3}+
  e^{\lambda k_{3,0}}\vec{p'}) ~.
 \label{npcons}
\end{equation}
While energy conservation is still ordinary also for non-planar diagrams,
momentum conservation is modified and it is modified in a way that cannot
even be described as a modified conservation law: the terms involving the
loop/integration momentum $\vec{k_3}$ do not cancel each other out in the
$\delta^{(3)}$ of (\ref{npcons}). It is easy to verify that these
non-planar contributions, while not implementing exactly the ordinary
energy-momentum conservation, are still mainly centered around ordinary
energy-momentum conservation (assuming reasonably good behaviour at
infinity of the expression under the integral). There is therefore some
deviation from ordinary conservation of energy-momentum, a sort of fuzzy
conservation of momentum, but it is plausible that the full theory
(whose construction
will also require the measure that we are here treating as an unknown)
would only predict a very small ($\lambda$-suppressed) deviation from
ordinary conservation of energy-momentum, possibly consistent with
observational limits. This dynamical issue is beyond the scopes of our
present study, focusing on kinematics, and we postpone it to future
studies.

Our next, and final, task is the study of the tree-level vertex.
The $O(\lambda)$ contribution to the four-point Green function can
be expressed in terms of $\bar{Z}^{(1)}$ through
\begin{equation}
G^{(4)}_{\lambda}(p_1,p_2,\dot{-}p_3,\dot{-}p_4)=
 \frac{\delta^4 \bar{Z}^{(1)}[J(k)]}{\delta J(\dot{-}p_1)\delta
 J(\dot{-}p_2)\delta J(p_3)\delta J(p_4)}\Bigg |_{J=0}
~,
\label{fourpf}
\end{equation}
and the tree-level vertex, sum of connected graphs contributing to
$G^{(4)}_{\lambda}(p_1,p_2,\dot{-}p_3,\dot{-}p_4)$, turns out
to have the form
\begin{eqnarray}
G^{(4)}_{\lambda}(p_1,p_2,\dot{-}p_3,\dot{-}p_4)_{connected} \!\!&=&\!\!
\frac{i\lambda}{24}
\int \left( \prod_{l=1}^4 \frac{d^4k_l}{2\pi}\xi(k_{l,0})\right)
\delta^{(4)}\left(\dot{\sum}_{k_1,k_2,k_3,k_4}\right)
~~~~~~~\nonumber\\
&~& \!\!\!\!\!\!\!\!\!\!\!\!\!\!\!\!\!\!\!\!\!\!\!\!\!\!\!\!\!\!\!
\!\!\!\!\!\!\!\!\!\!\!\!\!\!\!\!\!\!\!\!\!\!\!\!\!\!\!\!\!\!\!
\!\!\!\!\!\!\!\!\!\!\!\!\!\!\!\!\!\!\!
\Bigg( \frac{\delta^2
\bar{Z}^0(J(k))}{\delta J(\dot{-}p_{1})\delta J(\dot{-}k_1)}\Bigg |_{J=0}
\frac{\delta^2
\bar{Z}^0(J(k))}{\delta J(\dot{-}p_{2})\delta J(\dot{-}k_2)}\Bigg |_{J=0}
\frac{\delta^2
\bar{Z}^0(J(k))}{\delta J(p_3)\delta J(\dot{-}k_3)} \Bigg |_{J=0}
\frac{\delta^2
\bar{Z}^0(J(k))}{\delta J(p_4)\delta J(\dot{-}k_4)} \Bigg |_{J=0}
\nonumber\\
+ {\mathcal P}_{k_1,k_2,k_3,k_4} \Bigg)   ~.
\label{fourpf2}
\end{eqnarray}
From (\ref{fourpf2}) one can easily obtain more explicit formulas for the
tree-level vertex
$G^{(4)}_{tree}(p_1,p_2,\dot{-}p_3,\dot{-}p_4)_{connected}$.  As mentioned
we are primarily interested in establishing what are  the conservation
rules implemented at the vertex and how they are related to the
nonsymmetric structure of the coproduct. In this respect it is  important
to observe that each of the 24 terms generated by  the permutations
of $k_1,k_2,k_3,k_4$  is characterized by a different
conservation rule. This is completely  different from the behaviour of the
$\kappa \rightarrow \infty$ limit  (the limit in which our non-commutative
space-time turns into the  ordinary commutative Minkowski space-time),   in
which all permutations ${\mathcal P}_{k_1,k_2,k_3,k_4}$  lead to the same
conservation rule $\delta^{(4)}(-p_1-p_2+p_3+p_4)$.  In order to render
more explicit these comments on the faith of  energy-momentum conservation
in the vertices of field theories  on $\kappa$-Minkowski space-time,  it is
sufficient to observe that the term written out explicitly   in
Eq.~(\ref{fourpf2}) (the other 23
contributions are implicitly  introduced
through the permutations ${\mathcal P}_{k_1,k_2,k_3,k_4}$)  can be
rewritten as
\begin{eqnarray}
& & \!\!\!\! \frac{i\lambda}{24}
\int
\left(\prod_{l=1}^2 \frac{d^4k_l}{2\pi}\xi(k_{l,0})
\mu(p_{l,0})\mu(-p_{l,0})
 \frac{1}{2} \frac{\delta^{(4)}(p_l\dot{+}k_l)+
 \delta^{(4)}(k_l\dot{+}p_l)}  {{\mathcal C}_{\kappa}(p_l)-m^2}\right)
\label{fourpf3}\\ & &
\!\!\!\!\!\!\!\!\!\!\!  \left(\prod_{m=3}^4 \frac{d^4k_m}{2\pi}\xi(k_{m,0})
 \mu(p_{m,0}) \mu(-p_{m,0})\frac{1}{2}
\frac{\delta^{(4)}(\dot{-}p_m\dot{+}k_m)+
 \delta^{(4)}(k_m\dot{+}(\dot{-}p_m))} {{\mathcal
C}_{\kappa}(p_m)-m^2}\right)
\delta^{(4)}\left(\dot{\sum}_{k_1,k_2,k_3,k_4}\right) ~.
\nonumber
\end{eqnarray}
Then doing the $k_1,k_2,k_3,k_4$
integrations over the internal four-momenta it is easy to
establish that this contribution to the tree-level vertex enforces the
conservation rule
\begin{equation}
\delta^{(4)}(\dot{-}p_1 \dot{+}(\dot{-}p_2)\dot{+}p_3\dot{+}p_4)
~,
\label{deltafourpf}
\end{equation}
which corresponds to ordinary energy conservation,
$-p_{1,0}-p_{2,0}+p_{3,0}+p_{4,0}=0$, but  enforces a non-trivial and
non-symmetric rule of conservation of 3-momenta:
$-e^{\frac{-p_{1,0}}{\kappa}}\vec{p}_1-e^{-\frac{1}{\kappa}
(p_{1,0}+p_{2,0})}\vec{p}_2+e^{-\frac{1}{\kappa}(p_{1,0}+p_{2,0})}
\vec{p}_3 +e^{\frac{1}{\kappa}(-p_{1,0}-p_{2,0}+p_{3,0})} \vec{p}_4=0$.

The lack of symmetry of (\ref{deltafourpf}) is just of the type feared
in previous preliminary analyses~\cite{lukiFT,gacmaj} of field theory
in $\kappa$-Minkowski space-time, and was responsible for some
skepticism toward the physical applicability of such field theories
(since, of course, a physically acceptable theory should
describe collision processes in a way that treats symmetrically,
{\it e.g.}, two identical incoming particles).
However, in our approach to field theory
in $\kappa$-Minkowski space-time (\ref{fourpf3}) is just one of 24
contributions to the vertex (\ref{fourpf2}).
The other 23 terms are characterized
by permutations of the conservation rule (\ref{deltafourpf}),
which span over all possible ways to order the (deformed) sum
of $p_3$, $p_4$, $(\dot{-}p_1)$, and  $(\dot{-} p_2)$.
The overall structure of our interaction vertex is fully
symmetric\footnote{The fact that some sort of symmetrization
of the coproduct should emerge in the formalism had been conjectured
in Ref.~\cite{gacmaj}, where the puzzles implied by a naive
implementation of the non-symmetric coproduct were analyzed in detail.
Our result provides the correct realization of this conjectured symmetrization
(and the structure of our result is significantly different
from the simpler ansatzae considered in Ref.~\cite{gacmaj}).}
under exchanges of the momenta that enter it, thereby fulfilling
the condition for physical applicability of our field theory.
Our analysis shows that the new ingredient introduced by the $\kappa$
deformation is not the loss of particle-exchange symmetry, feared in
previous studies,
but rather a revision of the concept of energy-momentum
conservation for scattering processes:
since our vertex is not characterized by an
overall $\delta$-function, but instead it is split up into
24 pieces
each with its own different $\delta$-function, in a given scattering
process, with incoming particles characterized by four-momenta $p_1$
and $p_2$, it becomes impossible to predict the sum of the 3-momenta of
the outgoing particles. The theory only predicts that one of
our 24 energy-momentum-conservation rules must be satisfied and
assigns (equal) probabilities to each of these 24 channels.

These properties of vertices in $\kappa$-Minkowski space-time
represent a rather significant departure from conventional physics,
but they do make sense physically (identical particles
are treated symmetrically), and we are therefore providing a
key tool for testing whether Nature makes use of $\kappa$-Minkowski.
Making the reasonable assumption~\cite{dsr1dsr2}
that $\kappa$ should be of the order of the Planck scale
one easily checks that our prediction for new (non-)conservation
rules at the vertex is consistent with all available low-energy data.
(In the limit $p_0/\kappa \ll 1$ the 24 different conservation
rules that characterize our $\kappa$-deformed
vertex collapse into a single, and ordinary,
conservation rule.)
There is however one experimental context
in which our $\kappa$-deformed vertex could have observably large
consequences: astrophysics observations sensitive to the value of
the kinematic threshold for certain particle-production processes.
It is of encouragement for our proposal
that observations of ultra-high-energy cosmic rays~\cite{refCR}
and of Markarian501 photons~\cite{refMK} have recently obtained data
that appear to be in conflict with conventional theories and
appear to require~\cite{colgla,kifu,ita,aus,gactp1e2,gacqm100}
a deformation of the kinematic conservation rules applied to
collision processes.
$\kappa$-Minkowski space-time has been considered~\cite{ita,gactp1e2}
as a possible source of such a deformation of kinematics, but
the previously conjectured lack of symmetry of the vertex did not
allow direct comparison with data (and a simple-minded ansatz for the
symmetrization of the coproduct was shown not to lead to interesting
results~\cite{gactp1e2}).
Our analysis renders now possible this comparison with data,
and the fact that the data appear to be in conflict with the
conventional structure of vertices
encourages us to hope that the $\kappa$-deformed vertex might play a role
in the solution of the experimental paradox,
but we postpone this delicate phenomenological analysis
to a future study.

We close with some remarks summarizing the results here obtained.
We took off from some preliminary attempts~\cite{lukiFT,gacmaj}
to construct field theories in $\kappa$-Minkowski space-time.
Those previous studies had identified some intruiging features,
which might render such theories attractive as possible tools in
quantum-gravity research, but had also led to the alarming
suspect that the highly non-trivial structure of the $\kappa$-Poincar\'{e}
coproduct might render these theories unacceptable for physical
applications.
Our approach was mostly inspired by the previous study reported in
Ref.~\cite{gacmaj}, but we proposed that the correct starting point
for deriving
$\kappa$-Minkowski Green functions must be the corresponding generating
functional, just like in commutative spacetimes.
In the familiar commutative spacetimes one can easily guess the Green
functions from the structure of the action, and the previous attempts of
construction of field theory in $\kappa$-Minkowski space-time
relied on the assumption that very similar guess work could be adopted
in Lie-algebra noncommutative spacetimes; however, this is clearly
not the case. In particular, previous studies found that
the interaction vertices would not enjoy symmetry under exchange
of identical incoming particles~\cite{lukiFT,gacmaj},
while from our more fundamental generating-functional starting
point no such loss of symmetry was encountered.
Actually, at tree level the whole theory appears to be very intuitive,
without any dramatic departures from the structure of field theory
in commutative spacetimes.
Our tree-level results are sufficient (at least in principle, but the
required phenomenological analysis
appears to be rather challenging) for testing $\kappa$-Minkowski
through comparison with the recent exciting results of observations
of ultra-high-energy cosmic rays and of Markarian501 photons.

Beyond tree level the nature of the departures from the conventional
classical spacetime picture appears to become rather dramatic
in $\kappa$-Minkowski, at least from the field-theory perspective
here adopted. Evidence of such departures is found in
the non-planar-diagrams sector of our
one-loop analysis of the self-energy.
Whereas the theory at tree level enforces a standard concept
of energy-momentum conservation (although in an appropriately
adapted sense, reflecting the quantum, rather than classical,
symmetries of $\kappa$-Minkowski), beyond tree level energy-momentum
conservation is enforced only in a ``fuzzy" way: energy-momentum
is conserved on average (no preferred direction of the violations)
but violations of energy-momentum conservation are to be aspected
in any given particle-propagation processes.
Within our purely kinematical analysis (which allowed us to set aside
a delicate issue concerning the choice of measure for integration
over energy-momentum variables) we are of course not in the position
to estimate the quantitative significance of this effect,
but clearly there is strong motivation for future studies
to focus on this issue since it might lead to significant effects
(it is even possible that the anomalous effects would
be so large to be in conflict with available data, thereby
ruling out applications of $\kappa$-Minkowski in spacetime physics).
The issue of possible departures from energy-momentum conservation
for particle-propagation processes had already emerged in a previous
attempt~\cite{lukiFT}
to construct field theories in $\kappa$-Minkowski space-time.
However, within the approach adopted in Ref.~\cite{lukiFT}
this effect could not be studied consistently since the limited starting
point there adopted (guessing Feynman rules directly from an action,
rather than deriving them from a generating functional of Green functions)
did not give rise to the full structure of interaction
vertices\footnote{For example, the four-point vertex constructed in
Ref.~\cite{lukiFT} corresponds to only 1 of the 24 non-equivalent
permutations of external momenta that characterize the vertex we
constructed from the generating functional (and actually, even if one
wanted to choose 1 among the 24, it is difficult
to imagine an argument that could single out that particular one).
In addition, the approach developed in Ref.~\cite{lukiFT}
fails to uncover the key role played by the difference between
planar and nonplanar diagrams, which instead was imposed on us
by our constructive procedure based on the generating functional
of Green functions.
These issues concerning the role played by the differences between
planar and nonplanar diagrams appear to be
a general feature~\cite{seibwitt,suss,chong}
of field theory in noncommutative spacetimes, and it is encouraging
that our approach based on the generating functional of Green functions
proved to be consistent with this general expectation.}.

Although we focused here on a specific example of Lie-algebra non-commutative
space-time, $\kappa$-Minkowski, it appears likely that
the techniques we developed
be applicable also to field theories in
other Lie-algebra non-commutative
space-times, which are all affected by similar problems
associated with the highly non-trivial structure of the coproduct.

\baselineskip 12pt plus .5pt minus .5pt

\end{document}